\documentclass[reprint,amsmath,amssymb,aps,floatfix,showpacs,longbibliography,
prl,
]{revtex4-1}

\usepackage[utf8]{inputenc} 
\usepackage[babel]{csquotes} 

\usepackage{graphicx}
\usepackage{dcolumn}
\usepackage{bm}
\usepackage[]{units}	
\usepackage{xcolor}
\usepackage{soul}
\usepackage{changes}
\usepackage{hyperref}
\usepackage[all]{hypcap}
\usepackage{textcomp}
\usepackage{upgreek}
\usepackage{enumerate}
\usepackage{multirow} 

\newcommand{\ket}[1]{\ensuremath{\left|#1\right\rangle}}
\newcommand{\expval}[1]{\ensuremath{\left\langle#1\right\rangle}}

\newcommand{\reffig}[1]{Fig.~\ref{#1}}

\begin{document}

\title{High-resolution \enquote*{magic}-field spectroscopy on trapped polyatomic molecules}

\author{Alexander Prehn}
\author{Martin Ibr\"ugger}
\author{Gerhard Rempe}
\author{Martin Zeppenfeld}
	\email{Martin.Zeppenfeld@mpq.mpg.de}
\affiliation{Max-Planck-Institut f\"ur Quantenoptik, Hans-Kopfermann-Strasse 1, 85748 Garching, Germany}


\begin{abstract} 
Rapid progress in cooling and trapping of molecules has enabled first experiments on high resolution spectroscopy of trapped diatomic molecules, promising unprecedented precision. Extending this work to polyatomic molecules provides unique opportunities due to more complex geometries and additional internal degrees of freedom. Here, this is achieved by combining a homogeneous-field microstructured electric trap, rotational transitions with minimal Stark broadening at a \enquote*{magic} offset electric field, and optoelectrical Sisyphus cooling of molecules to the low millikelvin temperature regime. We thereby reduce Stark broadening on the $J=5\leftarrow4$ ($K=3$) transition of formaldehyde at 364\,GHz to well below 1\,kHz, observe Doppler-limited linewidths down to 3.8\,kHz, and determine the \enquote*{magic}-field line position with an uncertainty below $100\,$Hz. Our approach opens a multitude of possibilities for investigating diverse polyatomic molecule species.
\end{abstract}


\maketitle


Precise measurements of atomic transition frequencies, now reaching relative accuracies below $10^{-18}$~\cite{Brewer2019,Oelker2019}, have enabled groundbreaking progress in science and technology, ranging from optical clocks~\cite{Ludlow2015} to tests of fundamental physical theories~\cite{Griffith2009}. Despite the lower accuracy achieved with spectroscopic investigations of molecules so far, the structure and symmetry of molecular systems can often provide a more sensitive probe of fundamental physics~\cite{Steimle2014,DeMille2017}. A prominent example is the search for a permanent electric dipole moment of the electron~\cite{Cairncross2017,ACME2018}. Molecules are also used to test parity violation~\cite{Quack2011} and to measure fundamental constants~\cite{Mejri2015,Biesheuvel2016} and their possible time variation~\cite{Shelkovnikov2008,Truppe2013}. Moreover, precise knowledge of molecular constants is required in cold chemistry and collision studies~\cite{Bell2009,Bohn2017} as well as for the interpretation of astrophysical spectra~\cite{Bergin2007}.

In the past, precision spectroscopy on neutral molecules has typically been performed on beams of cold particles, limiting interrogation times to a few milliseconds~\cite{Veldhoven2004,Hudson2006a,Truppe2013,Baron2014,DeNijs2014,Cahn2014}. Performing spectroscopy on trapped molecules can provide substantial benefits, including increased interaction times and the ability to probe the same molecule repeatedly to investigate weak transitions. Here, the development of a variety of techniques to generate ensembles of molecules at sub-millikelvin temperatures~\cite{Ni2008, Quintero-Perez2014, Prehn2016, Norrgard2016} provides new perspectives. Already a number of groups have observed millisecond or longer timescale coherence with Ramsey spectroscopy in optical dipole or magnetic traps on a variety of diatomic molecule species, including KRb~\cite{Park2017}, NaK~\cite{Seesselberg2018}, Sr$_2$~\cite{Kondov2019}, and CaF~\cite{Caldwell2020}.

Extending similar methods to polyatomic molecules is of great interest. Symmetric-top and near-symmetric-top energy level structures provide closely spaced parity doublets with applications to quantum simulation~\cite{Wall2015}, quantum information processing (QIP)~\cite{Yu2019}, and precision measurements~\cite{Kozyryev2017}. The internal vibrational degrees of freedom of large molecules have been proposed for single molecule QIP~\cite{Tesch2002}. Many molecules found in astrophysical sources possess three or more atoms~\cite{Herbst2009}, with formaldehyde, the molecule investigated here, being of particular interest~\cite{Muller2017}. Polyatomic molecules are necessarily required for investigation of parity violation in chiral molecules~\cite{Quack2008}. All these applications rely on or would greatly benefit from precisely controlling molecular transition frequencies in a trap environment.



\begin{figure}[b]
	\centering
	\includegraphics{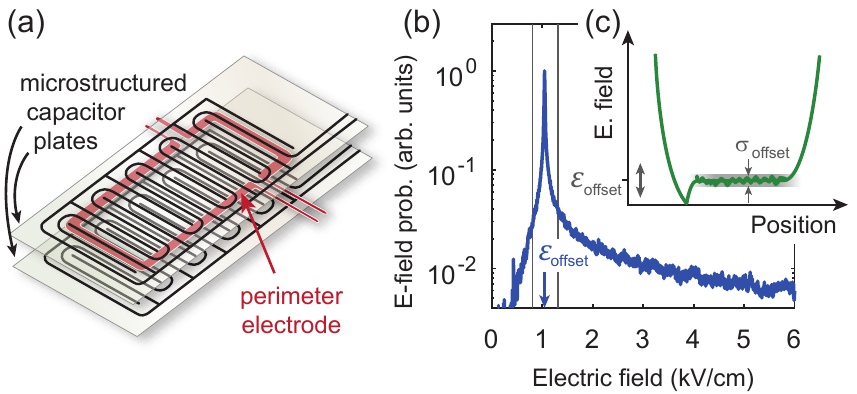}
	\caption{
		(a)~Illustration of the microstructured electric trap (not to scale).
		(b)~Electric field distribution in the trap, based on a 3d electrostatic field simulation. The vertical lines mark the electric field range of the inset in~\reffig{fig:Stark}(b).
		(c)~Simplified sketch of the position-dependent electric fields.
	}
	\label{fig:trap}
\end{figure}

In this paper, we advance the ability to perform few-kHz resolution spectroscopy on trapped molecules to include polyatomic species. This is achieved based on three main ingredients. First, molecules are trapped in a microstructured electric trap providing tunable homogeneous electric fields in a large fraction of the trapping region to reduce Stark broadening. Second, Stark broadening is further reduced by using suitable molecule transitions with vanishing variation in the differential Stark shift at a \enquote*{magic} value of the applied electric field. Third, Doppler broadening is reduced by cooling molecules to the low millikelvin temperature regime. We demonstrate this approach on formaldehyde, cooled via optoelectrical Sisyphus cooling. Stark broadening is suppressed to a sub-kHz level, allowing us to measure a spectrum with a FWHM linewidth of $3.8$\,kHz, dominated by Doppler broadening. We model the observed lineshape based on measured electric field distributions and energy distributions of molecules inside the trap, allowing us to closely reproduce the observed lineshape. This allows us to determine the line position of the \enquote*{magic}-field transition to within less than $100$\,Hz based on a statistical and systematic error analysis.

The first ingredient for our high-resolution spectroscopy is the use of our microstructured electric trap~\cite{Zeppenfeld2009,Englert2011}. The key elements for this trap design are a pair of capacitor plates holding an array of parallel electrodes and a perimeter electrode as sketched in \reffig{fig:trap}(a). This arrangement produces strong electric fields at the boundaries of the trapping volume ($\sim50\mathrm{kV/cm}$, corresponding to a nominal trap depth of $\sim1\,\mathrm{K}$) and a tunable offset electric field $\mathcal{E}_\mathrm{offset}$ in the trap center, resulting in a three-dimensional boxlike potential for molecules in low-field-seeking states. Stark broadening is caused by residual variations in the field strength in the trap center, with the homogeneity characterized by a roughness $\sigma_\mathrm{offset}$. The electric field distribution inside the trap for the voltage configuration used for the main measurements in this paper are shown in~\reffig{fig:trap}(b), and the main features of the trapping potential are visualized in~\reffig{fig:trap}(c). The trap design as well as its integration into the experimental apparatus and performance have been described earlier~\cite{Zeppenfeld2013,Englert2011}. Cooled molecules can be stored in the trap with 1/e decay times of up to one minute~\cite{Prehn2016}. 

To further suppress Stark broadening, the second ingredient for our high-resolution spectroscopy is to perform spectroscopy on pairs of states for which the differential Stark shift (meaning the difference in Stark shift between the two states) has an extremum (maximum or minimum) versus electric field. By setting the offset field to this value and for a sufficiently flat extremum, this can further suppress Stark broadening by many orders of magnitude. The offset electric field must be non-zero to avoid Majorana losses, and the pair of molecular states must be low-field-seeking to provide trapping.

\begin{figure}[t]
	\centering
	\includegraphics[width=0.235\textwidth]{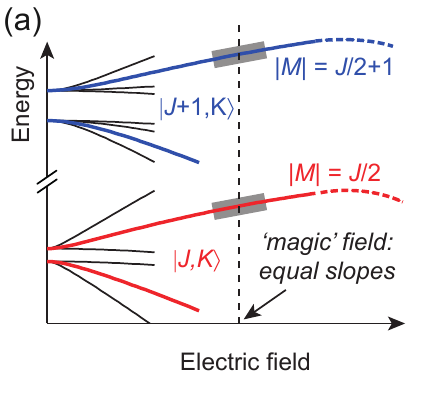}
	\includegraphics[width=0.235\textwidth]{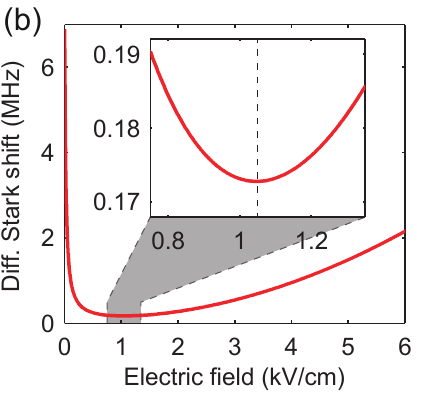}
	\caption{
		Illustration of concepts used for \enquote*{magic}-field spectroscopy.
		(a)~A very flat extremum in the differential Stark shift between a pair of states with the same first-order Stark shift is achieved due to the competition of splittings at very low electric field and higher order Stark shifts at very high electric fields.
		(b)~Differential Stark shift of the transition $\ket{4,3,2}\leftrightarrow\ket{5,3,3}$ of formaldehyde vs. electric field. The minimum is marked with a dashed line in the inset.
	}
	\label{fig:Stark}
\end{figure}

While pairs of states with an extremum in the differential Stark shift are present in all polar molecules, a generic scheme exists to identify such pairs of states for a (near-)symmetric-top energy level structure. This is specifically the case when corrections to the symmetric top energy level structure, e.g. due to hyperfine structure or rotational asymmetry, are small compared to the rotational spacing. In this case, linear Stark shifts extend over a large range in electric field, with the average near-constant orientation of the molecule symmetry axis with respect to the electric field direction given by $\expval{\cos\theta}=K\,M/J(J+1)$ for usual symmetric-top rotational quantum numbers $J, K, M$. The orientation and thus the slope of the linear Stark shift is equal for states $\ket{J,K,M}$ and $\ket{J+1,K,M+1}$ with $J=|2M|$ for arbitrary values of $K$ and $M$, resulting in small variation in the differential Stark shift over a large range in electric field. An extremum in the differential Stark shift is reached due to the competition between two effects, as illustrated in \reffig{fig:Stark}(a). For low electric fields, additional splittings, e.g. due to hyperfine structure and in the case of formaldehyde due to $K$-type doubling, must first be overcome before the Stark shift becomes linear. For large electric fields, coupling between neighboring rotational states results in higher order Stark shifts.

The final ingredient for our high-resolution spectroscopy is to reduce Doppler broadening. In our case, this is achieved by cooling molecules via optoelectrical Sisyphus cooling~\cite{Zeppenfeld2009,Zeppenfeld2012,Prehn2016}. In this cooling scheme, molecules are optically pumped to strongly trapped states in weak electric fields inside a trap and are coupled to more weakly trapped states in strong fields in the trap. Molecules thereby lose more kinetic energy while traveling from low electric fields to high electric fields than they regain while traveling back to low electric fields, resulting in an overall reduction in energy. This cooling scheme has been demonstrated for methyl fluoride and formaldehyde~\cite{Zeppenfeld2012,Prehn2016}, resulting in temperatures as low as $420\,\mu$K for an ensemble of about $300,000$ formaldehyde molecules. Since the scheme does not rely on fast electronic cycling transitions but also works for molecules with sufficiently strong vibrational transitions, it can be applied to an extremely broad range of molecules.

We demonstrate \enquote*{magic}-field spectroscopy on the $\ket{J,\pm K,\mp M}=\ket{4,3,2}\leftrightarrow\ket{5,3,3}$ transition of formaldehyde. This is the most easily accessible low-differential-Stark-shift transition from the state $\ket{3,3,3}$ in which molecules are prepared after cooling. The differential Stark shift for this transition is shown in \reffig{fig:Stark}(b), with a minimum at a field of about $\mathcal{E}_\mathrm{offset}=1.05\,\mathrm{kV/cm}$. A $1.1$\,\% variation about this field minimum, corresponding to the HWHM of the field distribution in our trap, causes a $20$\,Hz variation in the differential Stark shift. In contrast, the overall Stark shift of both states at this field is about $370$\,MHz, demonstrating the degree to which this choice of states reduces the differential Stark shift.

\begin{figure}[t]
	\centering
	\includegraphics{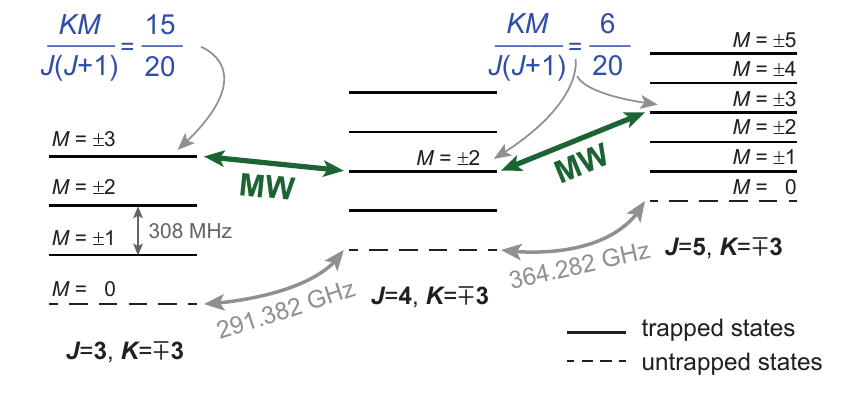}
	\caption{
		Level scheme for spectroscopy on the $\ket{4,3,2}\leftrightarrow\ket{5,3,3}$ transition.
	}
	\label{fig:scheme}
\end{figure}

An experiment proceeds as follows. Molecules are loaded into the electric trap from a velocity filtered source and cooled via optoelectrical Sisyphus cooling. The final molecule temperature can be varied by changing the cooling duration. Molecules are then prepared in the state $\ket{3,3,3}$ via optical pumping~\cite{Glockner2015a,Prehn2016}. Spectroscopy is performed via the level scheme schown in \reffig{fig:scheme}. After setting the offset field in the trap to the \enquote*{magic} field value, a pair of microwave fields coupling the states $\ket{3,3,3}$, $\ket{4,3,2}$, and $\ket{5,3,3}$ is applied for a duration of $500\,$ms. Specifically, radiation at approximately $291\,$GHz and $364\,$GHz is applied alternatingly for $4\,$ms and $6\,$ms, respectively. The exact frequency of the $364\,$GHz radiation is varied between individual shots of the experiment. This transfers population to the state $\ket{5,3,3}$ depending on the strength of the $\ket{4,3,2}\leftrightarrow\ket{5,3,3}$ transition at the given frequency. The intensity of the $364\,$GHz radiation must be strongly attenuated 
due to the narrow linewidth of the transition. To measure the population transferred to the state $\ket{5,3,3}$, molecules in the states $J=3$ and $J=4$ are eliminated from the trap via microwave depletion as described in Ref.~\cite{Gloeckner2015}. The final population of molecules in the trap is determined by unloading the molecules from the trap for measurement with a quadrupole mass spectrometer.

\begin{figure*}[tb]
	\centering
	\includegraphics{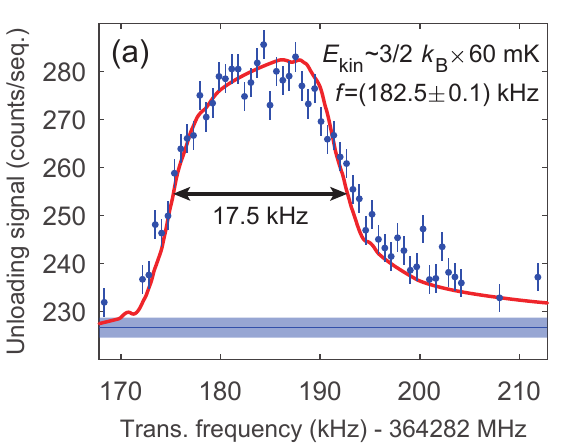}
	\includegraphics{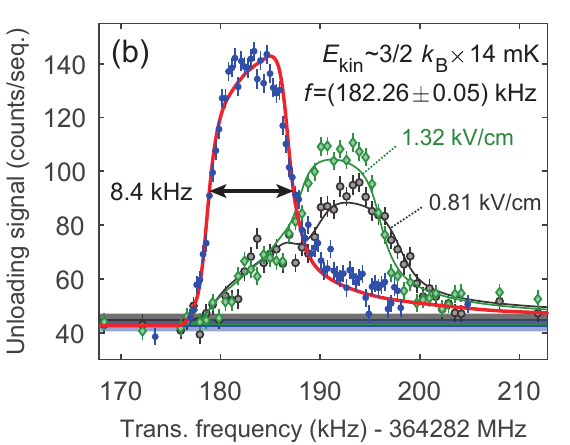}
	\includegraphics{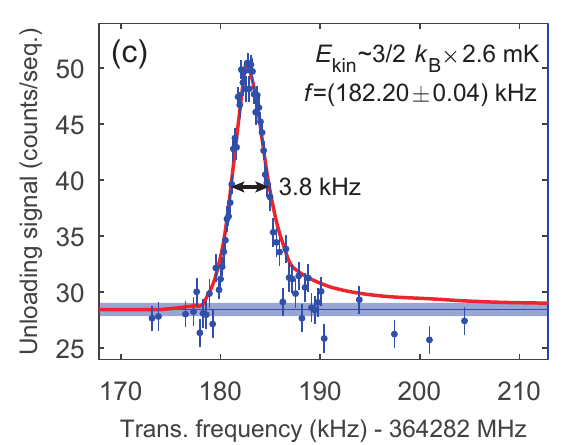}
	\caption{
		Measured and simulated spectra for various molecule kinetic energy. Two measurements are performed at a higher and lower electric offset field as indicated (center panel), all other measurements are performed at the \enquote*{magic} offset field in the trap. The FWHM and fitted \enquote*{magic}-field frequency for the \enquote*{magic}-field measurements are indicated. The horizontal lines mark the signals measured with the MW left off and determine the base lines of the modeled spectra. Error bars and shaded regions specify the 1\,$\sigma$ statistical error.
	}
	\label{fig:results}
\end{figure*}

Measured spectra for the $\ket{4,3,2}\leftrightarrow\ket{5,3,3}$ transition for different experimental parameters are shown in \reffig{fig:results}. In particular, we show spectra for three values of the molecular temperature with the offset field in the trap set to the \enquote*{magic} field of $1.05\,$kV/cm. Additionally, at the intermediate temperature, we show measurements for a higher and lower value of the offset field in the trap. The linewidth of the measured transition depends strongly on temperature, closely matching a $\sqrt{T}$ temperature dependence. This suggests Doppler broadening as the dominant broadening mechanism. For the lowest temperature, we achieve a FWHM linewidth of $3.8$\,kHz.

For a lower or higher electric offset field, the spectra broaden and are shifted to higher frequencies. This is consistent with the expected field dependence of the differential Stark shift shown in \reffig{fig:Stark}(b). Considering the change in the offset field is over $200\,$V/cm, this measurement directly demonstrates the low sensitivity of our transition to the electric field.

The key goal for any spectroscopy experiment is not just to obtain narrow lines to distinguish close-lying transitions, but also to accurately determine the center line position. In most spectroscopy experiments the desired line position is that which would be measured for an idealized system free from any interactions with the environment, since such a line position is well defined independent of any specific experiment. In our case, such a measurement is not possible, since the \enquote*{magic} field automatically introduces a large interaction with the environment. However, the line position precisely at the \enquote*{magic} electric field is equally well defined, and is thus an equally valid spectroscopic quantity.

To obtain the \enquote*{magic}-field line position, we model our lineshape based on our experimental parameters. It turns out that a fairly simple model only taking into account Doppler broadening and Stark broadening is sufficient for our purposes. While the simplification introduces some systematic error in the measured line position, this error is smaller than the statistical error and is included in the discussion on the systematic error of our measurement below. To model Doppler broadening, the energy distribution of molecules in the trap is measured via rf knife measurements as described in Ref.~\cite{Prehn2016}. Doppler broadening is calculated assuming an isotropic velocity distribution based on the energy distribution at the homogeneous field value in the trap. To model Stark broadening, either the theoretical electric field distribution shown in \reffig{fig:trap}(b) or a measured field distribution based on a non-\enquote*{magic} transition~\cite{Gloeckner2015} is combined with the theoretical differential Stark shift of our transition. The overall spectrum is obtained from convolution of the Doppler-broadened with the Stark-broadened spectrum.

The modeled lineshapes are shown together with the measured data in \reffig{fig:results}. The horizontal position and vertical scaling of the modeled lineshapes are determined from least-$\chi^2$ fitting to the experimental data, all other model parameters are determined from independent measurements. The model reproduces the observed lineshapes very well, demonstrating that Doppler broadening and Stark broadening are the dominant broadening mechanisms.

The horizontal fit of the model to the experimental data determines the \enquote*{magic}-field line position. The values obtained for the \enquote*{magic}-field measurements are indicated in \reffig{fig:results} together with the statistical uncertainty. These values are affected by a variety of sources of systematic error, and we estimate the magnitude of these errors for the lowest temperature measurement. First, the center of the electric field distribution in the trap might deviate from the \enquote*{magic}-field value by as much as $0.5\,\%$. This can shift the fitted \enquote*{magic}-field frequency by about $3.5\,$Hz. Second, the electric field distribution in the trap can be obtained both from simulations based on electrode geometry and from measurements on non-\enquote*{magic} transitions resulting in slightly different distributions. Switching between these distributions for the lineshape model shifts the fitted \enquote*{magic}-field frequency by about $34\,$Hz. Third, we model our lineshape by adding an offset and linear term as independent fit parameters. This shifts the fitted \enquote*{magic}-field frequency by about $23\,$Hz. Fourth, to estimate the effects of simplifications in our model, we construct a one-dimensional position-dependent electric field roughly matching the position dependent electric field along the narrow dimension of our trap. We determine the lineshape for molecules trapped in the resulting one dimensional potential both using our simplified lineshape model as well as for a full calculation based on the instantaneous Stark shift and Doppler shift for molecules propagating in this potential. This results in slightly different lineshapes with a frequency offset between the lineshapes based on least rms fitting of about $10\,$Hz. Finally, frequency calibration of our measurement is based on reference to a hydrogen maser with a relative accuracy of $3\times10^{-13}$, which thus does not relevantly contribute to our error. We estimate our systematic error as the root-sum-square of the indicated frequency shifts. This results in a value for the \enquote*{magic}-field frequency of $364,282,182,197\pm38_{\rm stat}\pm42_{\rm syst}\,$Hz based on the $2.6$\,mK measurement.

From the measured \enquote*{magic}-field transition frequency, the zero-field frequency can be calculated based on rotational constants of formaldehyde. We obtain $364,288,895.5\pm3\,$kHz for the $\ket{J,K_A;K_C}=\ket{4,3;1}\leftrightarrow\ket{5,3;2}$ transition, limited by the accuracy of rotational constants in the literature~\cite{Brunken2003}. This is consistent with the most accurate value of $364,288,884\pm30\,$kHz obtained previously for this transition~\cite{Cornet1980}.

In summary, we have demonstrated \enquote*{magic}-field spectroscopy on formaldehyde, obtaining a transition linewidth of $3.8\,$kHz and determining the center line position to less than $100\,$Hz. Future improvements will allow even narrower lines. In particular, the temperature of the molecules is currently limited by the large \enquote*{magic} electric offset field, which results in an increased roughness of the trapping potential and thereby limits the minimal temperature. Using different transitions with a lower \enquote*{magic} field value would allow the use of colder molecules with reduced Doppler broadening. For formaldehyde and similar molecules, this is systematically the case for \enquote*{magic}-field transitions with larger values of $K$. Various techniques from atomic physics could be used to circumvent Doppler broadening~\cite{Demtroeder15}.

Narrower lines would also be possible using a further improved~\cite{Zeppenfeld2013} or altered trap design. As an example, we consider electrically trapping formaldehyde in an electric field minimum using the present transition in fields varying by $0.1\,\%$ about the \enquote*{magic} field value. This would result in Stark broadening of about $200\,$mHz while providing a trap depth of about $40\,\mu$K$\times{\rm k_B}$.

Our approach is extremely general since \enquote*{magic} field transitions exist throughout the rotational, vibrational, and electronic spectrum of basically any polar molecule species. While the systematic approach we have discussed to identify such transitions in formaldehyde and similar molecules doesn't work in general, Stark shifts are always strongly nonlinear towards sufficiently high electric fields, offering many pairs of states with equal effective dipole moment at some field strength.

Measuring multiple \enquote*{magic}-field transitions in a given molecule species would allow a more precise determination of spectroscopic constants. Thus, the energy shift between the zero-field frequencies and the \enquote*{magic}-field frequencies predominantly depend on the rotational constants, with an additional small contribution at the present resolution due to the static polarisability. Fitting spectroscopic constants to a set of \enquote*{magic}-field frequencies or a set of zero-field frequencies for different transitions would thus be similarly valid. Accurate values for the zero-field frequencies could thereby also be obtained.

Our experimental technique has applications to other types of measurements. First, a reduction in the present linewidth by about a factor of $10$ would allow the hyperfine splitting of formaldehyde to be resolved. Second, the long interrogation time in our trap offers new possibilities for the investigation of very weak forbidden transitions. Third, the narrow transitions represent a major step towards coherent control over the internal molecular states. Here, the narrow linewidths observed indicate the existence of coherences on a long $\mu$s timescale. Demonstrating these explicitly would be possible, e.g., by replacing the weak, long microwave pulses used in the present experiment with pairs of short intense Ramsey pulses. However, this would also require illuminating the molecular ensemble with homogeneous intensity, which is not the case for the present experiments.

Two more longterm applications include observing parity violation in chiral molecules and placing more stringent limits on time variation of fundamental constants. The ability to circumvent Stark broadening to perform precision spectroscopy on trapped molecules is a major step for both applications. We emphasize that chiral symmetric top molecules such as phosphoric acid (H$_3$PO$_4$) are directly amenable to the same type of trapping and cooling techniques as are used here. Additional work is required to identify suitable molecular transitions.

\begin{acknowledgments}
Many thanks to Thomas Udem for helpful discussions. We acknowledged funding by the DFG via grant ZE 1096/2-1.
\end{acknowledgments}

\end{document}